%% file: polar.tex
\documentstyle[prl,aps,twocolumn]{revtex}

\input{psfig1}

\begin{document}
\twocolumn[
\hsize\textwidth\columnwidth\hsize\csname @twocolumnfalse\endcsname

\title{\hfill \\
A Hartree-Fock Study of Charge Redistribution in a
2D Mesoscopic Structure}
\author{A. Kambili}
\address{Department of Physics, University of Bath,
Claverton Down, Bath BA2 7AY, UK, e-mail: pysak@bath.ac.uk}
\author{H. Fehrmann}
\address{School of Physics and Chemistry, Lancaster University,
Lancaster LA1 4YB, UK}
\author{C. J. Lambert} 
\address{School of Physics and Chemistry, Lancaster University,
Lancaster LA1 4YB, UK}
\author{J. H. Jefferson} 
\address{Defence Evaluation and Research Agency, St. Andrews Road,
Malvern, Worcs. WR14 3PS, UK}

\date{\today} 

\maketitle
\begin{abstract}
In this paper, we investigate the ground state of two-dimensional
disordered cylinders which contain spinless, interacting electrons
using the Hartree-Fock approximation. Calculations of the
deviation of the polarization from uniformity reveal a tendency of
the charge to rearrange towards the ends of the system. The
presence of disorder results in fluctuations of the deviation
around its mean value, which are more pronounced when the disorder
strength is of the order of the interaction between the electrons.  
\end{abstract}

]

The existence of persistent currents in normal mesoscopic rings
threaded by a magnetic flux \cite{buttiker} has stimulated a
great deal of experimental and theoretical work, \cite{levy}-%
\cite{kato}. More recently, there have been studies of the ground
state of one-dimensional rings containing spinless fermions
(\cite{schmitteckert1}-\cite{schmitteckert3}) which take
into account both electron-electron interactions and disorder.
Strongly disordered rings with short-range interactions and
a half-filled band exhibit a reorganization of the charge in
the ground state. The charge density changes from an inhomogeneous
configuration due to the presence of the strong disorder to a
periodic array of charges as a result of the interactions.

In the present paper, we study the ground state of two-%
dimensional cylinders which contain spinless, interacting elect%
rons via Coulomb interactions, by solving self-consistently the
Hartree-Fock equation. The validity of the method has already been
established \cite{kambili} by studying the ground state of one-%
dimensional disordered rings and finding good agreement with the
exact calculations \cite{schmitteckert3}. Our aim is to examine
reorganization of the charge in the ground state, and in particular,
the manner in which charge tends to polarize towards the ends of
the cylinder, thereby producing spontaneous dipole or quadrapole
moments.

The system under investigation is a two-dimensional cylinder,
formed by a tight-binding lattice with $M$ sites in the longitu%
dinal direction ($x$) and $L$ sites in the transverse direction
($y$). We take periodic boundary conditions along the longitudinal
direction and free boundary conditions along the transverse. The
cylinder contains $N$ spinless electrons described by the Hamil%
tonian

\begin{equation}
H=\sum_{l=1}^{L\cdot M}\varepsilon_{l}\hat{c}^{\dagger}_{l}%
\hat{c}_{l}+\sum_{l,k=1}^{L\cdot M}V_{lk}\hat{c}^{\dagger}_{l}%
\hat{c}_{k}+\frac{1}{2}\sum_{l,k=1}^{L\cdot M}U_{lk}%
\hat{c}^{\dagger}_{l}\hat{c}_{l}\hat{c}^{\dagger}_{k}\hat{c}_{k}
\label{hamilt}
\end{equation}
where each site $l$ has coordinates $l=(x,y)$. The operators
$\hat{c}^{\dagger}_{l}$, $\hat{c}_{l}$ create and destroy a
particle at site $l$, respectively. $V_{lk}$ is the hopping
element between different sites. In the following, we will
restrict ourselves only to nearest-neighbour hopping elements
of strength $V_{lk}=-V$. $\varepsilon_{l}$ is the on-site energy
which is equal to $\varepsilon_{l}=4V+r_{l}$, where $r_{l}$ are
random numbers uniformly distributed over the range $[-W/2,
+W/2]$. $W$ is the strength of the disorder, and in the clean
case $W=0$. $U_{lk}$ is the interaction between the particles,
which has been taken to be long range,

\begin{equation}
U_{lk}=\frac{U}{|r_{1}-r_{2}|}
\end{equation}
where $r_{n}$ is the position of the $n$th particle. The Hartree-Fock
equation which corresponds to the Hamiltonian (\ref{hamilt}) is of
the form

\begin{eqnarray}
\varepsilon_{l}\Psi^{n}(l)-V\sum_{l'=n.n. of l}\Psi^{n}(l')+
\nonumber \\
\sum_{m=1}^{N}\sum_{k=1}^{L\cdot M}|\Psi^{m}(k)|^{2}U_{lk}%
\Psi^{n}(l)- \nonumber \\
\sum_{m=1}^{N}\sum_{k=1}^{L\cdot M}\Psi^{*m}(k)\Psi^{m}(l)U_{lk}
\Psi^{n}(k)= E_{n}\Psi^{n}(l)
\end{eqnarray}
where $\Psi^{n}(l)$ is the amplitude of the $n$th single-particle
wavefunction on site $l$, and $E_{n}$ is the corresponding single-%
particle energy. The third and fourth terms are the direct
and exchange potentials, respectively.

The probability of finding an electron on site ($l$) is
$\nu_{x,y}=\sum_{n=1}^{N}|\Psi^{n}(x,y)|^{2}$ and hence the mean
number of electrons per unit length in the transverse direction is

\begin{equation}
\rho(y)=\sum_{x=1}^{M}\nu_{x,y}
\end{equation}
We then define

\begin{equation}
p(y)=\frac{\rho(y)}{N}
\end{equation}
where $p(y)\Delta y$ is the probability of finding an electron in
the interval $[y,y+\Delta y]$. A measure of the deviation of the
charge from uniformity is the standard deviation of $p(y)$

\begin{equation}
\sigma^{2}=<y^{2}>-<y>^{2}
\end{equation}
where $<y^{n}>=\sum y^{n}p(y)\Delta y$. In the case of a clean
system ($W=0$) the electron density for a uniform distribution
must be $\rho(y)=N/L$, where $L$ is the system size in the
transverse direction (number of chains). This gives $\sigma^{2}%
/L^{2}=\frac{1}{12}$. On the other hand, for the extreme case
in which half the charge is at $L/2$ and the other half at $-L/2$,
$\rho(y)=\frac{N}{2}\delta (y+\frac{L}{2})+\frac{N}{2}\delta
(y-\frac{L}{2})$ which gives $\sigma^{2}/L^{2}=\frac{1}{4}$. Thus,
the length-normalized deviation of $p(y)$ can take values in the
interval $[1/12, 1/4]$, the minimum and the maximum values corres%
ponding to a uniform distribution and a maximally "polarized" one,
respectively.

\begin{figure}[h]
\centerline{\psfig{figure=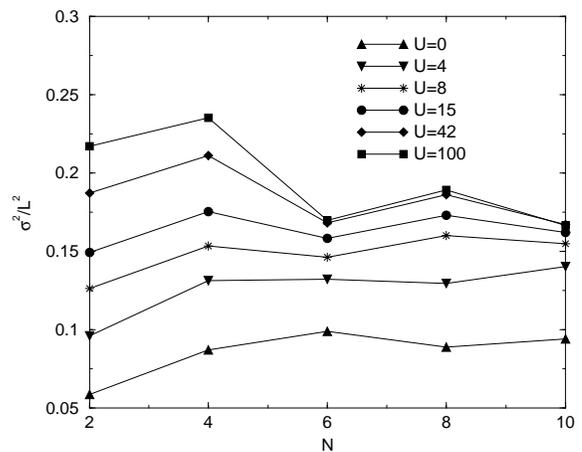,width=7.5cm}}
\caption{Length-normalized deviation of $p(y)$ with respect to the
number of particles $N$ for a clean cylinder $L=M=10$, for
different values of the interaction strength $U$.}
\label{1}
\end{figure}

In figure \ref{1} we present the normalized deviation with respect
to the number of particles $N$, for different values of $U$, and
$V=1$. The system has no disorder and its size is $L=M=10$. As we
increase the interactions between the particles, for small $N$,
$\sigma^{2}/L^{2}$ aproaches its maximum value, which indicates
total "polarization". This is the expected result since we are app%
roaching the electrostatic limit. When the system contains more
electrons there is still an increase in $\sigma^{2}/L^{2}$ as $U$
increases, but not all electrons move towards the ends of the cylinder.
We have confirmed that this again approaches the expected minimum
electrostatic energy configuration.

\begin{figure}[h]
\centerline{\psfig{figure=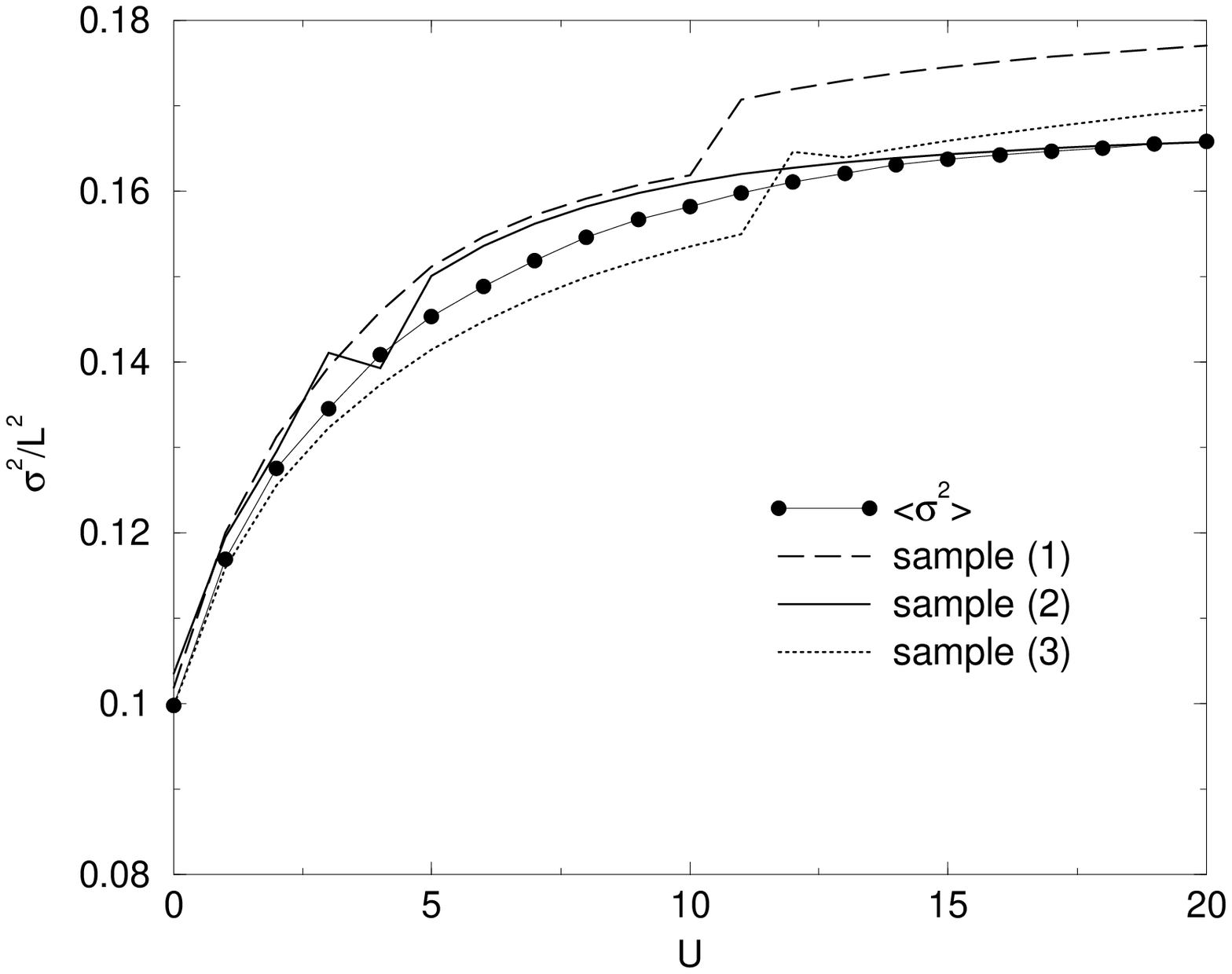,width=7.5cm}}
\caption{Length-normalized deviation of $p(y)$ with respect to the
interaction for disordered cylinders of size $L=M=10$ and $N=10$
electrons, with $W=2$. The dots represent the ensemble average of
$\sigma^{2}$ obtained from 200 samples. The different lines
represent individual disorder realizations.}
\label{2}
\end{figure}

\begin{figure}[h]
\centerline{\psfig{figure=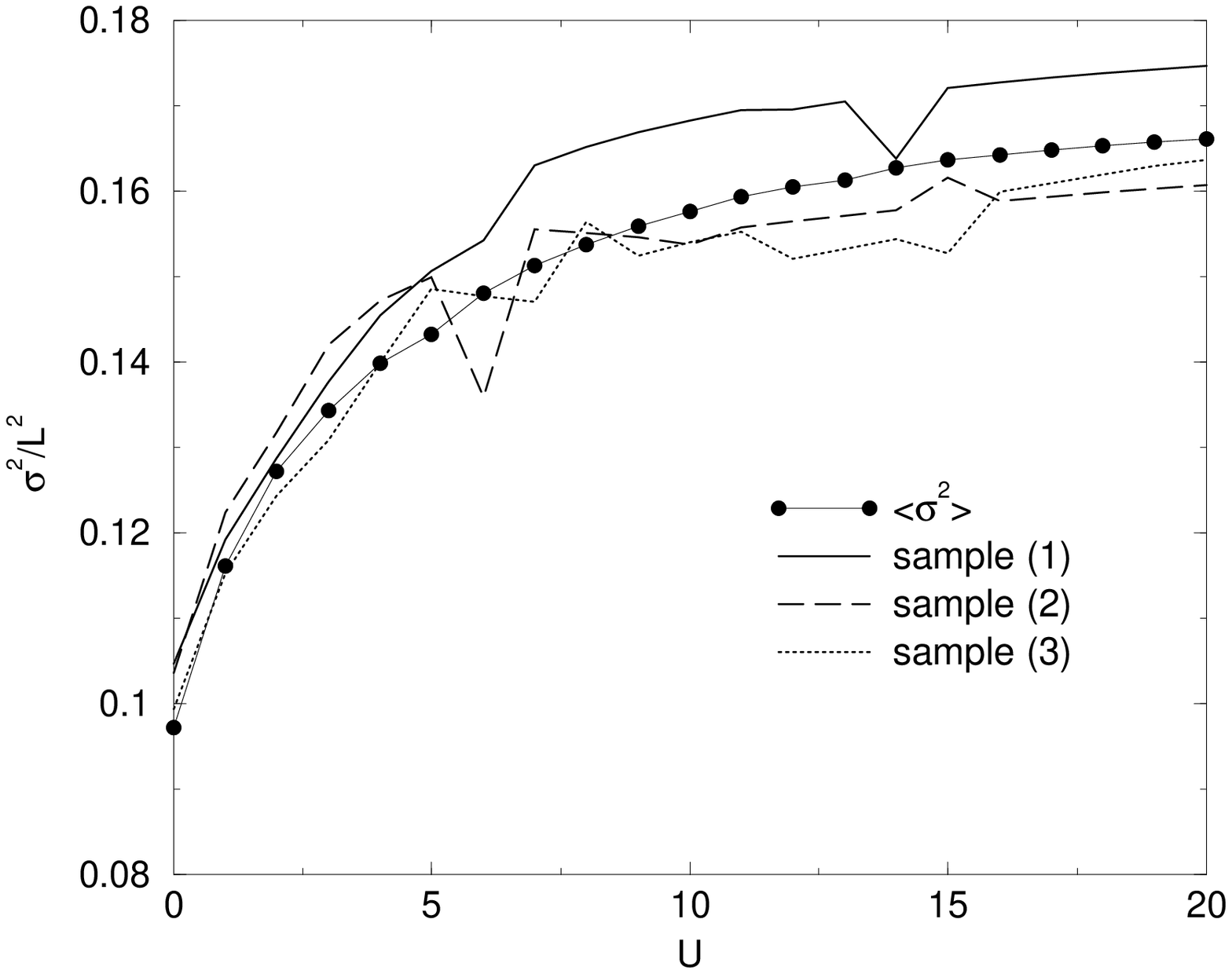,width=7.5cm}}
\caption{Length-normalized deviation of $p(y)$ with respect to
the interaction for disordered cylinders of size $L=M=10$ and
$N=10$ electrons, with $W=4$. The dots represent the ensemble
average of $\sigma^{2}$ obtained from 200 samples. The different
lines represent individual disorder realizations.}
\label{3}
\end{figure}

\begin{figure}[h]
\centerline{\psfig{figure=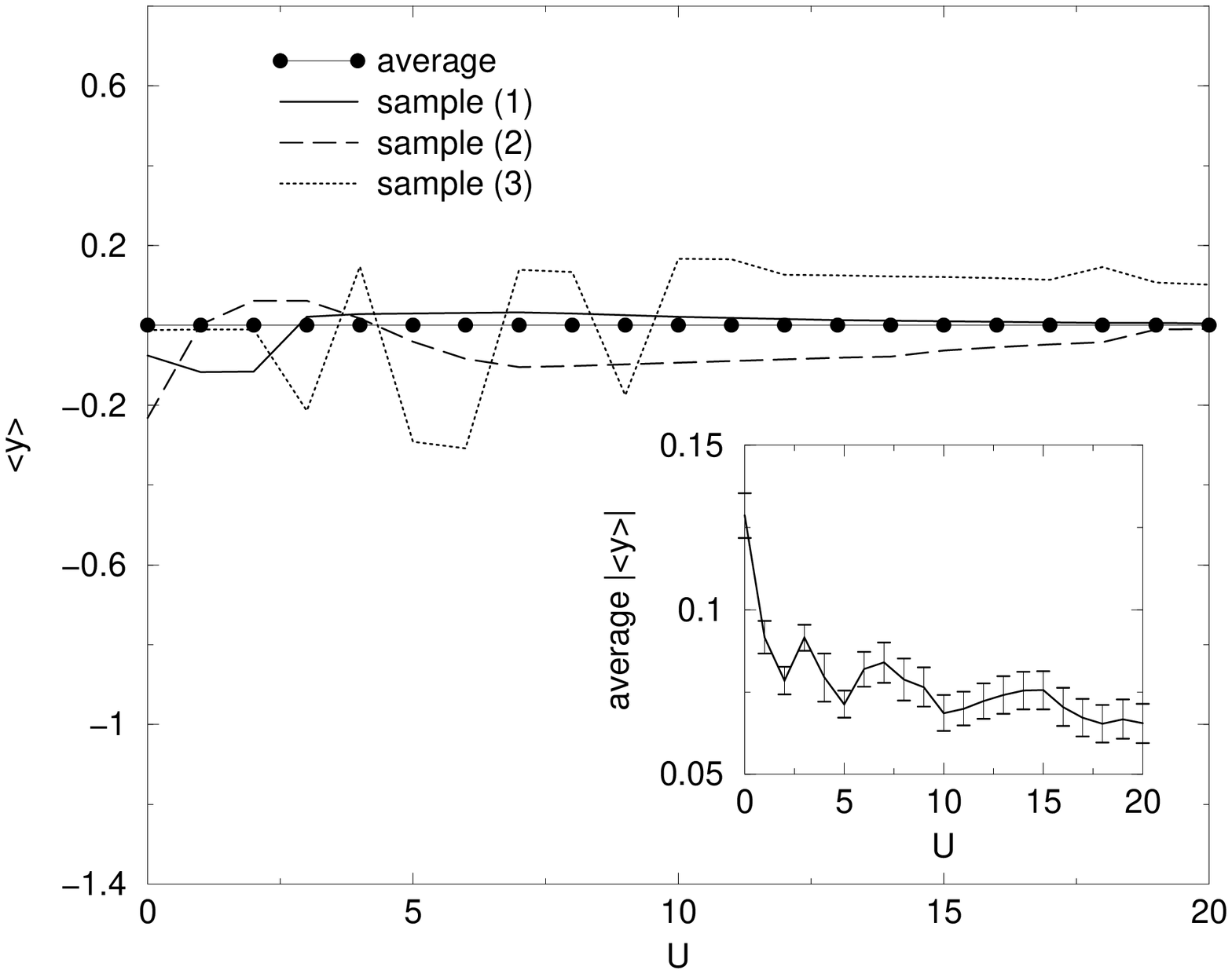,width=8cm}}
\caption{First moment of $p(y)$ versus $U$ for disordered cylinders
of size $L=M=10$ and $N=10$ electrons, with $W=2$. The dots
represent the ensemble average. The different lines represent
individual disorder realizations. The insert shows the average of
$|<y>|$ versus $U$.}
\label{4}
\end{figure}

\begin{figure}[h]
\centerline{\psfig{figure=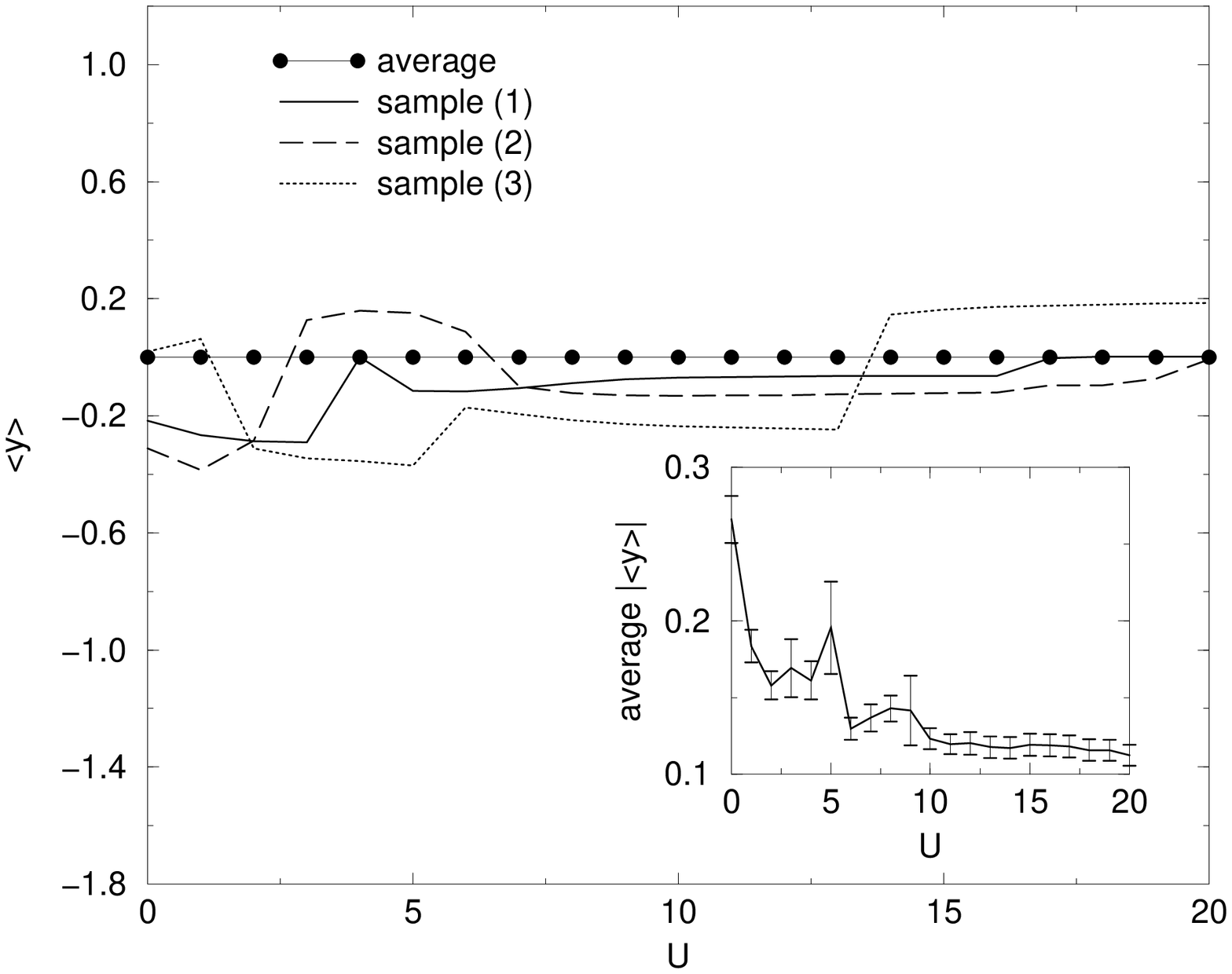,width=8cm}}
\caption{First moment of $p(y)$ versus $U$ for disordered cylinders
of size $L=M=10$ and $N=10$ electrons, with $W=4$. The dots
represent the ensemble average. The different lines represent
individual disorder realizations. The insert shows the average of
$|<y>|$ versus $U$.}
\label{5}
\end{figure}

In figures \ref{2} and \ref{3} we show $\sigma^{2}$ for a system
with non-zero disorder. In figure \ref{2} we plot the ensemble
average of $\sigma^{2}$ versus $U$, for disorder strength $W=2$
and $N=10$ along with $\sigma^{2}$ for three individual samples.
Individual samples show small fluctuations around the mean value
of $\sigma^{2}$ indicating that the charge tends to separate.
Increasing the disorder, as in figure \ref{3} where $W=4$, one
can see that the fluctuations increase with disorder, even though
the average (dots) remains practically unaffected by it. However, the
behaviour of the first moment $<y>$ for individual samples is more
pronounced than that for $\sigma^{2}$. In figures \ref{4} and
\ref{5} we have plotted $<y>$ for two different strengths of
disorder, $W=2$ and $W=4$, respectively. In the inserts of \ref{4}
and \ref{5}, the ensemble average of the absolute value of the
first moment is shown.

Figures \ref{2}, \ref{3}, \ref{4} and \ref{5} illustrate beha%
viour which we expect to be typical of small systems with free-end
boundary conditions in at least one direction. Namely that with
increasing $U$ the dipole moment $<y>$ induced by random potential
fluctuations (ie. $W\neq 0$) tends to zero, and that the charge
distribution becomes sharply peaked at the ends of the samples. A
key feature revealed by these figures is that at intermediate
values of $U$ (of order $W$), both $\sigma^{2}$ and $<y>$ exhibit
large sample-to-sample fluctuations about their means and large
fluctuations with increasing $U$, associated with charge
redistribution of the ground state. 

\begin{figure}[h]
\centerline{\psfig{figure=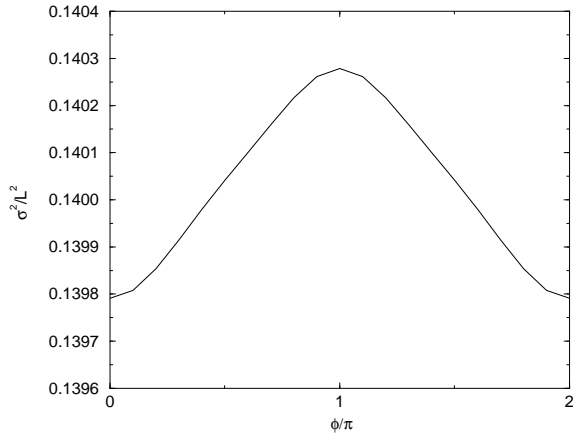,width=7.5cm}}
\caption{Length-normalized deviation of $p(y)$ for an individual
disorder realization, as a function of the magnetic field. The cylinder
is of size $L=M=10$ and contains $N=10$ electrons, with $W=2$.
The results have been obtained for $U=4$.}
\label{6}
\end{figure}

\begin{figure}[h]
\centerline{\psfig{figure=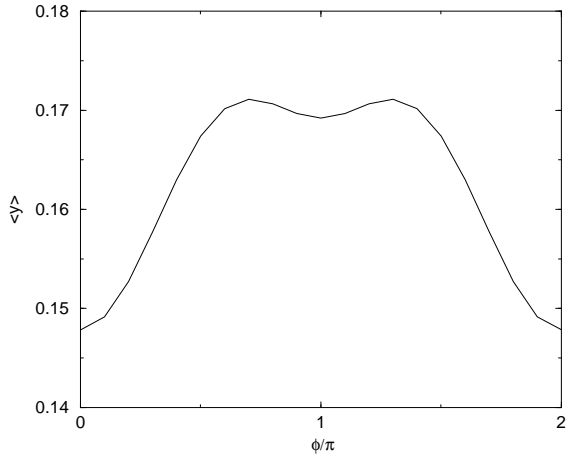,width=7.5cm}}
\caption{First moment of $p(y)$ for an individual disorder realization, with
respect the magnetic field. The cylinder is of size $L=M=10$ and contains
$N=10$ electrons, with $W=2$. The results have been obtained for $U=4$.}
\label{7}
\end{figure}

\begin{figure}[h]
\centerline{\psfig{figure=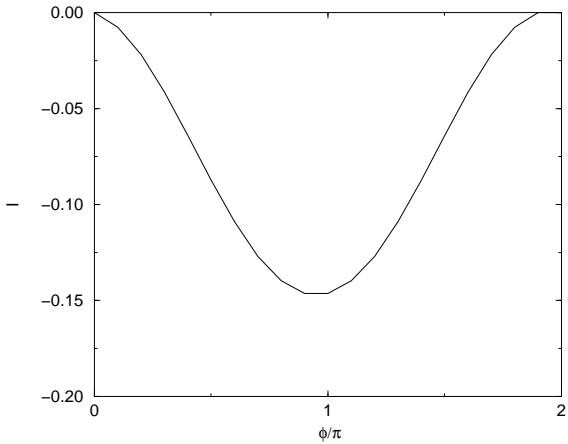,width=7.5cm}}
\caption{Current versus magnetic field for an individual disorder
realization. The cylinder is of size $L=M=10$ and contains
$N=10$ electrons, with $W=2$. The results have been obtained for $U=4$.}
\label{8}
\end{figure}

Finally, in figures \ref{6}, \ref{7} and \ref{8} we present
$\sigma^{2}$, $<y>$ and the current $I$, respectively, as a
function of the phase $\phi=2\pi \Phi/\Phi_{0}$, where $\Phi$ is
the magnetic flux threading the cylinder, and $\Phi_{0}$ is the
flux quantum. The current is equal to

\begin{equation}
I=-\frac{\delta E_{g}}{\delta \phi}
\end{equation}
where $E_{g}$ is the Hartree-Fock ground state energy. The results
show the behaviour of one disordered cylinder, for the case of
$W=2$ and for one value of the interaction strength ($U=4$) such
that it is of the order of the disorder strength. A common feature
in these three figures is that all quantities are symmetrical
around the value of the phase $\phi$ which corresponds to a
magnetic field of half a flux quantum. Such charge fluctuations
could  possibly be detected experimentally by placing a SET in the
vicinity of a sample, which couples to the electric field generated
by such a non-uniform charge distribution.

In this paper, we have made a Hartree-Fock study of the ground
state of two-dimensional cylinders which contain spinless,
interacting electrons. We found that the charge in the ground
state shows a separation towards the ends of the cylinder,
which is reflected in the polarization of the system. The
polarization shows fluctuations around its mean value in the
presence of disorder. These fluctuations are stronger in the first
moment of the charge disrtibution when the disorder, bandwidth
and interaction between the electrons are of the same order,
reflecting the competition between Mott and Anderson localization
(\cite{kramer}).

\end{document}

%% file: psfig1.tex
\def\PsfigVersion{1.9}
\ifx\undefined\psfig\else \fi

\let\LaTeXAtSign=\@
\let\@=\relax
\edef\psfigRestoreAt{\catcode`\@=\number\catcode`@\relax}
\catcode`\@=11\relax
\newwrite\@unused
\def\ps@typeout#1{{\let\protect\string\immediate\write\@unused{#1}}}
\ps@typeout{psfig/tex \PsfigVersion}

\def\figurepath{./}

\def\@nnil{\@nil}
\def\@empty{}
\def\@psdonoop#1\@@#2#3{}
\def\@psdo#1:=#2\do#3{\edef\@psdotmp{#2}\ifx\@psdotmp\@empty \else
    \expandafter\@psdoloop#2,\@nil,\@nil\@@#1{#3}\fi}
\def\@psdoloop#1,#2,#3\@@#4#5{\def#4{#1}\ifx #4\@nnil \else
       #5\def#4{#2}\ifx #4\@nnil \else#5\@ipsdoloop #3\@@#4{#5}\fi\fi}
\def\@ipsdoloop#1,#2\@@#3#4{\def#3{#1}\ifx #3\@nnil 
       \let\@nextwhile=\@psdonoop \else
      #4\relax\let\@nextwhile=\@ipsdoloop\fi\@nextwhile#2\@@#3{#4}}
\def\@tpsdo#1:=#2\do#3{\xdef\@psdotmp{#2}\ifx\@psdotmp\@empty \else
    \@tpsdoloop#2\@nil\@nil\@@#1{#3}\fi}
\def\@tpsdoloop#1#2\@@#3#4{\def#3{#1}\ifx #3\@nnil 
       \let\@nextwhile=\@psdonoop \else
      #4\relax\let\@nextwhile=\@tpsdoloop\fi\@nextwhile#2\@@#3{#4}}
\ifx\undefined\fbox
\newdimen\fboxrule
\newdimen\fboxsep
\newdimen\ps@tempdima
\newbox\ps@tempboxa
\long\def\fbox#1{\leavevmode\setbox\ps@tempboxa\hbox{#1}\ps@tempdima\fboxrule
    \advance\ps@tempdima \fboxsep \advance\ps@tempdima \dp\ps@tempboxa
   \hbox{\lower \ps@tempdima\hbox
  {\vbox{\hrule height \fboxrule
          \hbox{\vrule width \fboxrule \hskip\fboxsep
          \vbox{\vskip\fboxsep \box\ps@tempboxa\vskip\fboxsep}\hskip 
                 \fboxsep\vrule width \fboxrule}
                 \hrule height \fboxrule}}}}
\fi
\newread\ps@stream
\newif\ifnot@eof       
\newif\if@noisy        
\newif\if@atend        
\newif\if@psfile       
{\catcode`\%=12\global\gdef\epsf@start{
\def\epsf@PS{PS}
\def\epsf@getbb#1{%
\openin\ps@stream=#1
\ifeof\ps@stream\ps@typeout{Error, File #1 not found}\else
   {\not@eoftrue \chardef\other=12
    \def\do##1{\catcode`##1=\other}\dospecials \catcode`\ =10
    \loop
       \if@psfile
	  \read\ps@stream to \epsf@fileline
       \else{
	  \obeyspaces
          \read\ps@stream to \epsf@tmp\global\let\epsf@fileline\epsf@tmp}
       \fi
       \ifeof\ps@stream\not@eoffalse\else
       \if@psfile\else
       \expandafter\epsf@test\epsf@fileline:. \\%
       \fi
          \expandafter\epsf@aux\epsf@fileline:. \\%
       \fi
   \ifnot@eof\repeat
   }\closein\ps@stream\fi}%
\long\def\epsf@test#1#2#3:#4\\{\def\epsf@testit{#1#2}
			\ifx\epsf@testit\epsf@start\else
\ps@typeout{Warning! File does not start with `\epsf@start'.  It may not be a PostScript file.}
			\fi
			\@psfiletrue} 
{\catcode`\%=12\global\let\epsf@percent=
\long\def\epsf@aux#1#2:#3\\{\ifx#1\epsf@percent
   \def\epsf@testit{#2}\ifx\epsf@testit\epsf@bblit
	\@atendfalse
        \epsf@atend #3 . \\%
	\if@atend	
	   \if@verbose{
		\ps@typeout{psfig: found `(atend)'; continuing search}
	   }\fi
        \else
        \epsf@grab #3 . . . \\%
        \not@eoffalse
        \global\no@bbfalse
        \fi
   \fi\fi}%
\def\epsf@grab #1 #2 #3 #4 #5\\{%
   \global\def\epsf@llx{#1}\ifx\epsf@llx\empty
      \epsf@grab #2 #3 #4 #5 .\\\else
   \global\def\epsf@lly{#2}%
   \global\def\epsf@urx{#3}\global\def\epsf@ury{#4}\fi}%
\def\epsf@atendlit{(atend)} 
\def\epsf@atend #1 #2 #3\\{%
   \def\epsf@tmp{#1}\ifx\epsf@tmp\empty
      \epsf@atend #2 #3 .\\\else
   \ifx\epsf@tmp\epsf@atendlit\@atendtrue\fi\fi}

\chardef\psletter = 11 
\chardef\other = 12

\newif \ifdebug 
\newif\ifc@mpute 
\c@mputetrue 

\let\then = \relax
\def\r@dian{pt }
\let\r@dians = \r@dian
\let\dimensionless@nit = \r@dian
\let\dimensionless@nits = \dimensionless@nit
\def\internal@nit{sp }
\let\internal@nits = \internal@nit
\newif\ifstillc@nverging
\def \Mess@ge #1{\ifdebug \then \message {#1} \fi}

{ 
	\catcode `\@ = \psletter
	\gdef \nodimen {\expandafter \n@dimen \the \dimen}
	\gdef \term #1 #2 #3%
	       {\edef \t@ {\the #1}
		\edef \t@@ {\expandafter \n@dimen \the #2\r@dian}%
		\t@rm {\t@} {\t@@} {#3}%
	       }
	\gdef \t@rm #1 #2 #3%
	       {{%
		\count 0 = 0
		\dimen 0 = 1 \dimensionless@nit
		\dimen 2 = #2\relax
		\Mess@ge {Calculating term #1 of \nodimen 2}%
		\loop
		\ifnum	\count 0 < #1
		\then	\advance \count 0 by 1
			\Mess@ge {Iteration \the \count 0 \space}%
			\Multiply \dimen 0 by {\dimen 2}%
			\Mess@ge {After multiplication, term = \nodimen 0}%
			\Divide \dimen 0 by {\count 0}%
			\Mess@ge {After division, term = \nodimen 0}%
		\repeat
		\Mess@ge {Final value for term #1 of 
				\nodimen 2 \space is \nodimen 0}%
		\xdef \Term {#3 = \nodimen 0 \r@dians}%
		\aftergroup \Term
	       }}
	\catcode `\p = \other
	\catcode `\t = \other
	\gdef \n@dimen #1pt{#1} 
}

\def \Divide #1by #2{\divide #1 by #2} 

\def \Multiply #1by #2
       {{
	\count 0 = #1\relax
	\count 2 = #2\relax
	\count 4 = 65536
	\Mess@ge {Before scaling, count 0 = \the \count 0 \space and
			count 2 = \the \count 2}%
	\ifnum	\count 0 > 32767 
	\then	\divide \count 0 by 4
		\divide \count 4 by 4
	\else	\ifnum	\count 0 < -32767
		\then	\divide \count 0 by 4
			\divide \count 4 by 4
		\else
		\fi
	\fi
	\ifnum	\count 2 > 32767 
	\then	\divide \count 2 by 4
		\divide \count 4 by 4
	\else	\ifnum	\count 2 < -32767
		\then	\divide \count 2 by 4
			\divide \count 4 by 4
		\else
		\fi
	\fi
	\multiply \count 0 by \count 2
	\divide \count 0 by \count 4
	\xdef \product {#1 = \the \count 0 \internal@nits}%
	\aftergroup \product
       }}

\def\r@duce{\ifdim\dimen0 > 90\r@dian \then   
		\multiply\dimen0 by -1
		\advance\dimen0 by 180\r@dian
		\r@duce
	    \else \ifdim\dimen0 < -90\r@dian \then  
		\advance\dimen0 by 360\r@dian
		\r@duce
		\fi
	    \fi}

\def\Sine#1%
       {{%
	\dimen 0 = #1 \r@dian
	\r@duce
	\ifdim\dimen0 = -90\r@dian \then
	   \dimen4 = -1\r@dian
	   \c@mputefalse
	\fi
	\ifdim\dimen0 = 90\r@dian \then
	   \dimen4 = 1\r@dian
	   \c@mputefalse
	\fi
	\ifdim\dimen0 = 0\r@dian \then
	   \dimen4 = 0\r@dian
	   \c@mputefalse
	\fi
	\ifc@mpute \then
		\divide\dimen0 by 180
		\dimen0=3.141592654\dimen0
		\dimen 2 = 3.1415926535897963\r@dian 
		\divide\dimen 2 by 2 
		\Mess@ge {Sin: calculating Sin of \nodimen 0}%
		\count 0 = 1 
		\dimen 2 = 1 \r@dian 
		\dimen 4 = 0 \r@dian 
		\loop
			\ifnum	\dimen 2 = 0 
			\then	\stillc@nvergingfalse 
			\else	\stillc@nvergingtrue
			\fi
			\ifstillc@nverging 
			\then	\term {\count 0} {\dimen 0} {\dimen 2}%
				\advance \count 0 by 2
				\count 2 = \count 0
				\divide \count 2 by 2
				\ifodd	\count 2 
				\then	\advance \dimen 4 by \dimen 2
				\else	\advance \dimen 4 by -\dimen 2
				\fi
		\repeat
	\fi		
			\xdef \sine {\nodimen 4}%
       }}

\def\Cosine#1{\ifx\sine\UnDefined\edef\Savesine{\relax}\else
		             \edef\Savesine{\sine}\fi
	{\dimen0=#1\r@dian\advance\dimen0 by 90\r@dian
	 \Sine{\nodimen 0}
	 \xdef\cosine{\sine}
	 \xdef\sine{\Savesine}}}	      

\def\psdraft{
	\def\@psdraft{0}
}
\def\psfull{
	\def\@psdraft{100}
}

\psfull

\newif\if@scalefirst
\def\psscalefirst{\@scalefirsttrue}
\def\psrotatefirst{\@scalefirstfalse}
\psrotatefirst

\newif\if@draftbox
\def\psnodraftbox{
	\@draftboxfalse
}
\def\psdraftbox{
	\@draftboxtrue
}
\@draftboxtrue

\newif\if@prologfile
\newif\if@postlogfile
\def\pssilent{
	\@noisyfalse
}
\def\psnoisy{
	\@noisytrue
}
\psnoisy
\newif\if@bbllx
\newif\if@bblly
\newif\if@bburx
\newif\if@bbury
\newif\if@height
\newif\if@width
\newif\if@rheight
\newif\if@rwidth
\newif\if@angle
\newif\if@clip
\newif\if@verbose
\def\@p@@sclip#1{\@cliptrue}

\newif\if@decmpr

\def\@p@@sfigure#1{\def\@p@sfile{null}\def\@p@sbbfile{null}
	        \openin1=#1.bb
		\ifeof1\closein1
	        	\openin1=\figurepath#1.bb
			\ifeof1\closein1
			        \openin1=#1
				\ifeof1\closein1%
				       \openin1=\figurepath#1
					\ifeof1
					   \ps@typeout{Error, File #1 not found}
						\if@bbllx\if@bblly
				   		\if@bburx\if@bbury
			      				\def\@p@sfile{#1}%
			      				\def\@p@sbbfile{#1}%
							\@decmprfalse
				  	   	\fi\fi\fi\fi
					\else\closein1
				    		\def\@p@sfile{\figurepath#1}%
				    		\def\@p@sbbfile{\figurepath#1}%
						\@decmprfalse
	                       		\fi%
			 	\else\closein1%
					\def\@p@sfile{#1}
					\def\@p@sbbfile{#1}
					\@decmprfalse
			 	\fi
			\else
				\def\@p@sfile{\figurepath#1}
				\def\@p@sbbfile{\figurepath#1.bb}
				\@decmprtrue
			\fi
		\else
			\def\@p@sfile{#1}
			\def\@p@sbbfile{#1.bb}
			\@decmprtrue
		\fi}

\def\@p@@sfile#1{\@p@@sfigure{#1}}

\def\@p@@sbbllx#1{
		\@bbllxtrue
		\dimen100=#1
		\edef\@p@sbbllx{\number\dimen100}
}
\def\@p@@sbblly#1{
		\@bbllytrue
		\dimen100=#1
		\edef\@p@sbblly{\number\dimen100}
}
\def\@p@@sbburx#1{
		\@bburxtrue
		\dimen100=#1
		\edef\@p@sbburx{\number\dimen100}
}
\def\@p@@sbbury#1{
		\@bburytrue
		\dimen100=#1
		\edef\@p@sbbury{\number\dimen100}
}
\def\@p@@sheight#1{
		\@heighttrue
		\dimen100=#1
   		\edef\@p@sheight{\number\dimen100}
}
\def\@p@@swidth#1{
		\@widthtrue
		\dimen100=#1
		\edef\@p@swidth{\number\dimen100}
}
\def\@p@@srheight#1{
		\@rheighttrue
		\dimen100=#1
		\edef\@p@srheight{\number\dimen100}
}
\def\@p@@srwidth#1{
		\@rwidthtrue
		\dimen100=#1
		\edef\@p@srwidth{\number\dimen100}
}
\def\@p@@sangle#1{
		\@angletrue
		\edef\@p@sangle{#1} 
}
\def\@p@@ssilent#1{ 
		\@verbosefalse
}
\def\@p@@sprolog#1{\@prologfiletrue\def\@prologfileval{#1}}
\def\@p@@spostlog#1{\@postlogfiletrue\def\@postlogfileval{#1}}
\def\@cs@name#1{\csname #1\endcsname}
\def\@setparms#1=#2,{\@cs@name{@p@@s#1}{#2}}
\def\ps@init@parms{
		\@bbllxfalse \@bbllyfalse
		\@bburxfalse \@bburyfalse
		\@heightfalse \@widthfalse
		\@rheightfalse \@rwidthfalse
		\def\@p@sbbllx{}\def\@p@sbblly{}
		\def\@p@sbburx{}\def\@p@sbbury{}
		\def\@p@sheight{}\def\@p@swidth{}
		\def\@p@srheight{}\def\@p@srwidth{}
		\def\@p@sangle{0}
		\def\@p@sfile{} \def\@p@sbbfile{}
		\def\@p@scost{10}
		\def\@sc{}
		\@prologfilefalse
		\@postlogfilefalse
		\@clipfalse
		\if@noisy
			\@verbosetrue
		\else
			\@verbosefalse
		\fi
}
\def\parse@ps@parms#1{
	 	\@psdo\@psfiga:=#1\do
		   {\expandafter\@setparms\@psfiga,}}
\newif\ifno@bb
\def\bb@missing{
	\if@verbose{
		\ps@typeout{psfig: searching \@p@sbbfile \space  for bounding box}
	}\fi
	\no@bbtrue
	\epsf@getbb{\@p@sbbfile}
        \ifno@bb \else \bb@cull\epsf@llx\epsf@lly\epsf@urx\epsf@ury\fi
}	
\def\bb@cull#1#2#3#4{
	\dimen100=#1 bp\edef\@p@sbbllx{\number\dimen100}
	\dimen100=#2 bp\edef\@p@sbblly{\number\dimen100}
	\dimen100=#3 bp\edef\@p@sbburx{\number\dimen100}
	\dimen100=#4 bp\edef\@p@sbbury{\number\dimen100}
	\no@bbfalse
}
\newdimen\p@intvaluex
\newdimen\p@intvaluey
\def\rotate@#1#2{{\dimen0=#1 sp\dimen1=#2 sp
		  \global\p@intvaluex=\cosine\dimen0
		  \dimen3=\sine\dimen1
		  \global\advance\p@intvaluex by -\dimen3
		  \global\p@intvaluey=\sine\dimen0
		  \dimen3=\cosine\dimen1
		  \global\advance\p@intvaluey by \dimen3
		  }}
\def\compute@bb{
		\no@bbfalse
		\if@bbllx \else \no@bbtrue \fi
		\if@bblly \else \no@bbtrue \fi
		\if@bburx \else \no@bbtrue \fi
		\if@bbury \else \no@bbtrue \fi
		\ifno@bb \bb@missing \fi
		\ifno@bb \ps@typeout{FATAL ERROR: no bb supplied or found}
			\no-bb-error
		\fi
		\count203=\@p@sbburx
		\count204=\@p@sbbury
		\advance\count203 by -\@p@sbbllx
		\advance\count204 by -\@p@sbblly
		\edef\ps@bbw{\number\count203}
		\edef\ps@bbh{\number\count204}
		\if@angle 
			\Sine{\@p@sangle}\Cosine{\@p@sangle}
	        	{\dimen100=\maxdimen\xdef\r@p@sbbllx{\number\dimen100}
					    \xdef\r@p@sbblly{\number\dimen100}
			                    \xdef\r@p@sbburx{-\number\dimen100}
					    \xdef\r@p@sbbury{-\number\dimen100}}
                        \def\minmaxtest{
			   \ifnum\number\p@intvaluex<\r@p@sbbllx
			      \xdef\r@p@sbbllx{\number\p@intvaluex}\fi
			   \ifnum\number\p@intvaluex>\r@p@sbburx
			      \xdef\r@p@sbburx{\number\p@intvaluex}\fi
			   \ifnum\number\p@intvaluey<\r@p@sbblly
			      \xdef\r@p@sbblly{\number\p@intvaluey}\fi
			   \ifnum\number\p@intvaluey>\r@p@sbbury
			      \xdef\r@p@sbbury{\number\p@intvaluey}\fi
			   }
			\rotate@{\@p@sbbllx}{\@p@sbblly}
			\minmaxtest
			\rotate@{\@p@sbbllx}{\@p@sbbury}
			\minmaxtest
			\rotate@{\@p@sbburx}{\@p@sbblly}
			\minmaxtest
			\rotate@{\@p@sbburx}{\@p@sbbury}
			\minmaxtest
			\edef\@p@sbbllx{\r@p@sbbllx}\edef\@p@sbblly{\r@p@sbblly}
			\edef\@p@sbburx{\r@p@sbburx}\edef\@p@sbbury{\r@p@sbbury}
		\fi
		\count203=\@p@sbburx
		\count204=\@p@sbbury
		\advance\count203 by -\@p@sbbllx
		\advance\count204 by -\@p@sbblly
		\edef\@bbw{\number\count203}
		\edef\@bbh{\number\count204}
}
\def\in@hundreds#1#2#3{\count240=#2 \count241=#3
		     \count100=\count240	
		     \divide\count100 by \count241
		     \count101=\count100
		     \multiply\count101 by \count241
		     \advance\count240 by -\count101
		     \multiply\count240 by 10
		     \count101=\count240	
		     \divide\count101 by \count241
		     \count102=\count101
		     \multiply\count102 by \count241
		     \advance\count240 by -\count102
		     \multiply\count240 by 10
		     \count102=\count240	
		     \divide\count102 by \count241
		     \count200=#1\count205=0
		     \count201=\count200
			\multiply\count201 by \count100
		 	\advance\count205 by \count201
		     \count201=\count200
			\divide\count201 by 10
			\multiply\count201 by \count101
			\advance\count205 by \count201
		     \count201=\count200
			\divide\count201 by 100
			\multiply\count201 by \count102
			\advance\count205 by \count201
		     \edef\@result{\number\count205}
}
\def\compute@wfromh{
		\in@hundreds{\@p@sheight}{\@bbw}{\@bbh}
		\edef\@p@swidth{\@result}
}
\def\compute@hfromw{
	        \in@hundreds{\@p@swidth}{\@bbh}{\@bbw}
		\edef\@p@sheight{\@result}
}
\def\compute@handw{
		\if@height 
			\if@width
			\else
				\compute@wfromh
			\fi
		\else 
			\if@width
				\compute@hfromw
			\else
				\edef\@p@sheight{\@bbh}
				\edef\@p@swidth{\@bbw}
			\fi
		\fi
}
\def\compute@resv{
		\if@rheight \else \edef\@p@srheight{\@p@sheight} \fi
		\if@rwidth \else \edef\@p@srwidth{\@p@swidth} \fi
}
\def\compute@sizes{
	\compute@bb
	\if@scalefirst\if@angle
	\if@width
	   \in@hundreds{\@p@swidth}{\@bbw}{\ps@bbw}
	   \edef\@p@swidth{\@result}
	\fi
	\if@height
	   \in@hundreds{\@p@sheight}{\@bbh}{\ps@bbh}
	   \edef\@p@sheight{\@result}
	\fi
	\fi\fi
	\compute@handw
	\compute@resv}

\def\psfig#1{\vbox {
	\ps@init@parms
	\parse@ps@parms{#1}
	\compute@sizes
	\ifnum\@p@scost<\@psdraft{
		\special{ps::[begin] 	\@p@swidth \space \@p@sheight \space
				\@p@sbbllx \space \@p@sbblly \space
				\@p@sbburx \space \@p@sbbury \space
				startTexFig \space }
		\if@angle
			\special {ps:: \@p@sangle \space rotate \space} 
		\fi
		\if@clip{
			\if@verbose{
				\ps@typeout{(clip)}
			}\fi
			\special{ps:: doclip \space }
		}\fi
		\if@prologfile
		    \special{ps: plotfile \@prologfileval \space } \fi
		\if@decmpr{
			\if@verbose{
				\ps@typeout{psfig: including \@p@sfile.Z \space }
			}\fi
			\special{ps: plotfile "`zcat \@p@sfile.Z" \space }
		}\else{
			\if@verbose{
				\ps@typeout{psfig: including \@p@sfile \space }
			}\fi
			\special{ps: plotfile \@p@sfile \space }
		}\fi
		\if@postlogfile
		    \special{ps: plotfile \@postlogfileval \space } \fi
		\special{ps::[end] endTexFig \space }
		\vbox to \@p@srheight sp{
			\hbox to \@p@srwidth sp{
				\hss
			}
		\vss
		}
	}\else{
		\if@draftbox{		
			\hbox{\frame{\vbox to \@p@srheight sp{
			\vss
			\hbox to \@p@srwidth sp{ \hss \@p@sfile \hss }
			\vss
			}}}
		}\else{
			\vbox to \@p@srheight sp{
			\vss
			\hbox to \@p@srwidth sp{\hss}
			\vss
			}
		}\fi

	}\fi
}}
\psfigRestoreAt
\let\@=\LaTeXAtSign